\documentclass[a4paper, amsfonts, amssymb, amsmath, reprint, showkeys, nofootinbib, twoside]{revtex4-1}
\usepackage[english]{babel}
\usepackage[utf8]{inputenc}
\usepackage[colorinlistoftodos, color=green!40, prependcaption]{todonotes}
\usepackage{amsthm}
\usepackage{mathtools}
\usepackage{physics}
\usepackage{xcolor}
\usepackage{graphicx}
\usepackage[left=23mm,right=13mm,top=35mm,columnsep=15pt]{geometry} 
\usepackage{adjustbox}
\usepackage{placeins}
\usepackage[T1]{fontenc}
\usepackage{lipsum}
\usepackage{csquotes}
\usepackage{dcolumn}
\usepackage{bm}

\newcommand{\be}{\begin{equation}}       
\newcommand{\ee}{\end{equation}}       
\newcommand{\beqn}{\begin{eqnarray}}

\newcommand{\dpt}{\Delta{p_T^2}}
\newcommand{\CR}{C_{\mathcal{R}}}
\newcommand{\CA}{C_{\mathcal{A}}}
\newcommand{\CF}{C_{\mathcal{F}}}
\usepackage[pdftex, pdftitle={Article}, pdfauthor={Author}]{hyperref} 
\begin{document}
\title{Nuclear transverse momentum imbalance in the color dipole approach at the LHC regime}

\author{F. G. Ben$^1$}
\author{A. V. Giannini$^{2}$} 
\author{M. V. T. Machado$^1$}
\affiliation{$^1$ Universidade Federal do Rio Grande do Sul, Institute of Physics, Porto Alegre, RS, Brazil}
\affiliation{$^2$ Instituto de F\'isica Gleb Wataghin, Universidade Estadual de Campinas,
R. S\'ergio Buarque de Holanda, 777, 13083-859, Campinas, Brazil}


\begin{abstract}
Transverse momentum broadening of a parton propagating through a large nucleus is evaluated in the color dipole approach using different models for the dipole cross section or unintegrated gluon distribution, which lead to different values of the coefficient $C_{\mathcal{F}}(0,s)$. Numerical calculations are compared to data extracted from LHCb and ALICE experiments for nuclear broadening of $J/\psi$. We find that different models which describe the small-$x$ data predict values of $\dpt$ that agree reasonably well with experiment, specially for forward rapidity. The centrality dependence was also analysed and the models are consistent with the ALICE measurements.
\end{abstract}


\maketitle

\section{Introduction} \label{sec:outline}

High energy partons propagating through nuclear matter experience an increase in transverse momentum, due to multiple interactions with the medium. This broadening in transverse momentum is higher in proton-nucleus ($pA$) collisions than in proton-proton ($pp$) collisions, and the increase of the mean transverse momentum squared of the produced particle in $pA$ collisions, compared to $pp$ collisions, is defined as nuclear broadening:
\be 
\dpt = \langle p_T^2 \rangle _{pA} - \langle p_T^2 \rangle_{pp}.
\ee

Different theoretical approaches have been used to describe this broadening. Among them, one has the QCD color dipole approach,  Glauber multiple scattering \cite{Gyulassy:2002yv},
the Color Glass Condensate (CGC) framework \cite{Dumitru:2001jn,Rezaeian:2012ye} as well as the high-twist expansion of matrix elements\cite{Xing:2012ii,Kang:2016ron,Kang:2014hha}. In particular, in the color dipole approach \cite{dolejvsi1993colour,johnson2001broadening}, as well as in the BDMPS approach \cite{baier1997radiative}, the parton acquires transverse momentum through a random walk through nuclear matter, undergoing multiple rescatterings. In fact, as shown in Ref. \cite{raufeisen2003relating}, both descriptions are equivalent and are related to the higher twist factorization formalism, in which broadening arises from the exchange of a single soft gluon. Although all of these descriptions rely on nonperturbative inputs, the color dipole approach has the advantage of relying on a well developed phenomenology from deep-inelastic scattering \cite{johnson2001broadening}. The purpose of this Letter is to evaluate $\dpt$ in the color dipole approach using different phenomenological models as an input and compare its predictions. We also compare the numerical results to data extracted from the LHC for nuclear broadening of $J/\psi$.


Let us begin by briefly summarizing the main points from the dipole approach. Transverse momentum broadening of a high energy parton propagating through large nuclear matter is given by
\be
\dpt^\mathcal{R} = \langle T_A \rangle C_{\mathcal{R}}(0,s),
\label{eq:dpt}
\ee
where $\langle T_A \rangle = \int d^2b \, T_A^2 (b)/A$ is the nuclear thickness function averaged over impact parameter $b$. Assuming uniform nuclear density, $\langle T_A \rangle = 2\rho_A L$, where $\rho_A = 0.16$ fm$^{-3}$ is the nuclear density, $L = 3 R_A/4$, and $R_A$ is the nuclear radius. 
We follow the notation of Ref. \cite{raufeisen2003relating}, in which the index $\mathcal{R}$ represents the projectile parton, i.e., $\mathcal{R} = \mathcal{F}$ for a quark and $\mathcal{R}=\mathcal{A}$ for a gluon. The coefficient $\CR$ contains the nonperturbative physics and it arises from the expression for the total cross section $\sigma_{q\bar{q}} (r_T,s)$ for the interaction between a nucleon and a colorless dipole $q\bar{q}$ having transverse separation $r_T$ and c.m. energy squared $s$:
\beqn
\CF(0,s) &=& \frac{d}{dr_T^2} \sigma_{q\bar{q}} (r_T,s) \Bigr|_{r_T \rightarrow 0} \label{eq:cf}\\
\CA(0,s) &=& 9\CF(0,s)/4.
\end{eqnarray}
As shown in Ref. \cite{johnson2001broadening}, the effect of broadening increases with energy, and it will depend on the phenomenological model for $\sigma_{q\bar{q}} (r_T,s)$.


One can now look for different models for the dipole cross section and compare its predictions.
We start with the parametrization of Kopeliovich, Schäfer and Tarasov (KST) \cite{kopeliovich2000nonperturbative}, the one used in Ref. \cite{raufeisen2003relating}, which has a saturated form of the dipole cross section adjusted to low-$Q^2$ deep inelastic scattering (DIS) data:
\be
\sigma_{q\bar{q}}(r_T,s) = \sigma_0(s)\Bigr[ 1 - \exp \Bigr(- \frac{r_T^2}{R_0^2(s)} \Bigr) \Bigr],
\label{eq:KST}
\ee
in which an explicit energy dependence is introduced through
\be
\sigma_0(s) = \sigma^{\pi p}_{\mathrm{tot}}(s) \Bigr( 1 + \frac{3R_0^2(s)}{8\langle r_{ch}^2 \rangle _{\pi}} \Bigr),
\ee
where $\sigma^{\pi p}_{\mathrm{tot}}(s) = 23.6 \times (s/s_0)^{0.08}$, $\langle r_{ch}^2 \rangle _{\pi} = 0.44 \pm 0.01$ fm$^2$ is the mean-squared pion charge radius, and $R_0(s) = 0.88 \, \mathrm{fm}\times (s/s_0)^{-\lambda/2}$ with $\lambda = 0.28$ and $s_0 = 1000$ GeV$^2$ is the energy-dependent radius.
Using Eq. \eqref{eq:cf}, this leads to
\be
\CF^{\mathrm{KST}}(0,s) = \frac{\sigma_0(s)}{R_0^2(s)}.
\label{eq:cfKST}
\ee

In a different model, by Schildknecht, Surrow and Tentyukov (SST) \cite{schildknecht2001scaling,Schildknecht:2020oug}, we have
\be
\sigma_{q\bar{q}}(r_T,s) = \sigma^{(\infty)}\frac{1}{24}r_T^2 \Lambda ^2(s),
\label{eq:SST}
\ee 
where $\Lambda^2(s) = C_1(s + W_0^2)^{C_2}$, $\sigma^{(\infty)} = 48 \, \mathrm{GeV}^{-2} = 18.7 \, \mathrm{mb}$, $C_1 = 0.34 \pm 0.05$ and $C_2 = 0.27 \pm 0.01$. The SST model provides simple analytic expressions for proton structure function $F_2$, $F_L$, as well as for the photoabsorption cross section. It includes color transparency and saturation properties that are dependent on the relative magnitude of $Q^2$ and $W$ in the small-$x$ domain. The results are compared to a global analysis of all experimental data available, and agreement is quite good. Applying Eq. \eqref{eq:SST} to Eq. \eqref{eq:cf}, we find that
\be
\CF^{\mathrm{SST}}(0,s) = \frac{\sigma^{(\infty)}}{24}\Lambda^2(s).
\label{eq:cfSST}
\ee

In the model by Donnachie and Dosch (DD) \cite{donnachie2001diffractive,Donnachie:2001wt}, the dipole cross section is given by
\be
\sigma_d(r_T) = 0.098 (\langle g^2FF \rangle a^4)^2 r_T \Bigr[1 - \exp(-\frac{r_T}{3.1a}) \Bigr],
\label{eq:DD}
\ee
with $\sigma_{q\bar{q}}(r_T,s) = \sigma_d \times (\dfrac{s}{s_o})^{\epsilon}$. This model is based on the expectation value of two light-like Wilson loops and the dipole cross section is obtained from dipole-dipole interaction. The quantity $\langle g^2FF \rangle$ describes the gluon condensate  and the parameter $a$ is the correlation length of the two-gluon correlator. At small $r_T$, the hard Pomeron contribution dominates and then $\epsilon = \epsilon_{hard} = 0.42$. The dimensionless constant $\langle g^2FF \rangle a^4$ has the numerical value of $23.8$, $s_o = 20^2$ GeV$^2$ and $a = 0.346$ fm. These parameters were taken from lattice results and calibrated to high energy proton-proton scattering. If $a$ and $r_T$ are measured in fm, the result is in milibarn. At small $r_T$, it leads to
\be
\CF^{\mathrm{DD}}(0,s) = \frac{0.0098(\langle g^2FF \rangle a^4)^2 [\mathrm{fm}]}{3.1a} \Bigr( \frac{s}{s_o} \Bigr)^{0.42}.
\label{eq:cfDD}
\ee
The DD model is the basis for the posterior FKS model \cite{forshaw1999extracting,forshaw2000predicting,mcdermott2002colour}, a two-component model in which $\sigma(s,r) = \sigma_{\mathrm{soft}}(s,r)+ \sigma_{\mathrm{hard}}(s,r)$. Physics at small $r$ is dominated by the hard term
\be
\sigma_{\mathrm{hard}} = (a_2^Hr^2 + a_6^H r^6) \exp(-\nu_H r)(r^2 s)^{\lambda_H}.
\label{eq:FKS}
\ee
Due to more complex shape on $r$ for both hard and soft Pomeron contribution, FKS is not directly suitable for our analysis.

All phenomenological models for $\sigma_{q\bar{q}} (r_T,s)$ presented so far parametrize their energy evolution according to particular considerations that may or may not encode some features presented in the CGC. A more formal approach within the CGC framework (while still assuming a simplified description of the nuclear geometry and disregarding final state effects such as energy loss) is to make use of dipole models that are solutions of the running coupling Balitsky-Kovchegov (rcBK) equation~\cite{Kovchegov:2006vj,Balitsky:2006wa,Gardi:2006rp,Balitsky:2007feb}, which describes the energy/rapidity evolution of the dipole scattering amplitude in the fundamental representation, $\mathcal{N}_F(r_T,x)$. Such quantity is related to the dipole cross section simply by $\sigma_{q\bar{q}} =\sigma_0\, \mathcal{N}_F(r_T,x)$. Differently from previous models, $\sigma_0$ is assumed to be constant and all dynamics is encoded in $\mathcal{N}_F(r,x)$. 

Solving the rcBK equation requires an initial condition, that is, the value of $\mathcal{N}_F(r_T,x=x_0)$, where $x_0=0.01$; therefore, different initial conditions may lead to different dipole cross sections. In this work, we make use of rcBK evolved dipole cross sections constrained by HERA data provided by the AAMQS collaboration. In order to explore different energy dependencies that may arise from different initial conditions, we consider the ``GBW'' fit from \cite{Albacete:2009fh} and the fit ``h'' from~\cite{Albacete:2010sy}, which employs a modified McLerran-Venugopalan (MV) model initial condition that accounts for an anomalous dimension larger than unity. We verified that one gets essentially the same results by using the fit ``e'' from~\cite{Albacete:2010sy}, which has a marginally worse chi-squared per degree-of-freedom with respect to HERA data. Results obtained in this way will be identified by their initial condition, {\it i}.{\it e}. ``GBW'' and ``$\rm MV^\gamma$''.

As presented in Ref. \cite{johnson2001broadening}, the coefficient $\CF(0,s)$ may also be obtained directly from the unintegrated gluon density
\be
\CF(0,s) = \frac{\pi}{3}\int d^2k \frac{\alpha_s(k^2) k^2}{k^4}\mathcal{F}(x,k^2),
\label{eq:cfint}
\ee
where $\mathcal{F}(x,k^2) = \partial G(x,k^2)/ \partial (\ln k^2)$. We shall consider the recent model by Moriggi, Peccini and Machado (MPM) \cite{moriggi2020investigating}, in which $\mathcal{F}(x,k^2)$ is modeled as
\be
\mathcal{F}^{\mathrm{MPM}}(x,k^2) = \frac{3\sigma_o}{4\pi^2\alpha_s} \frac{1+\delta n}{Q_s^2}\frac{k^4}{(1+\tau)^{2+\delta n}},
\label{eq:MPM}
\ee
with $Q_s^2 = 1\, [\mathrm{GeV}^2]\times(x/xo)^{0.33}$ and $\delta n = a \tau^b$, where $\tau = k^2/Q_s^2$. The parameters $\sigma_0$, $x_o$, $a$ and $b$ were fitted against DIS data for $x < 0.01$ (see Ref. \cite{moriggi2020investigating} for details). This model is based on the geometric scaling scaling property and it is constructed in order to describe DIS data and $p_T$ spectra of produced hadrons at high energy proton-proton simultaneously. It has also been extended to proton-nucleus and nucleus-nucleus collisions in Ref. \cite{Moriggi:2020qla}. In order to calculate the energy dependence of $\CF(0,s)$ in this model we need to know the value of $x$. We shall follow the approach of Ref. \cite{johnson2001broadening} and use the minimal value of $x$ permitted by kinematics, $x = 4k^2/s$, in the integral in Eq. \eqref{eq:cfint}. The corresponding value of $\CF^{\mathrm{MPM}}(0,s)$ should be an upper bound in this model.
\begin{figure}[t]
\centering
\includegraphics[scale=0.6]{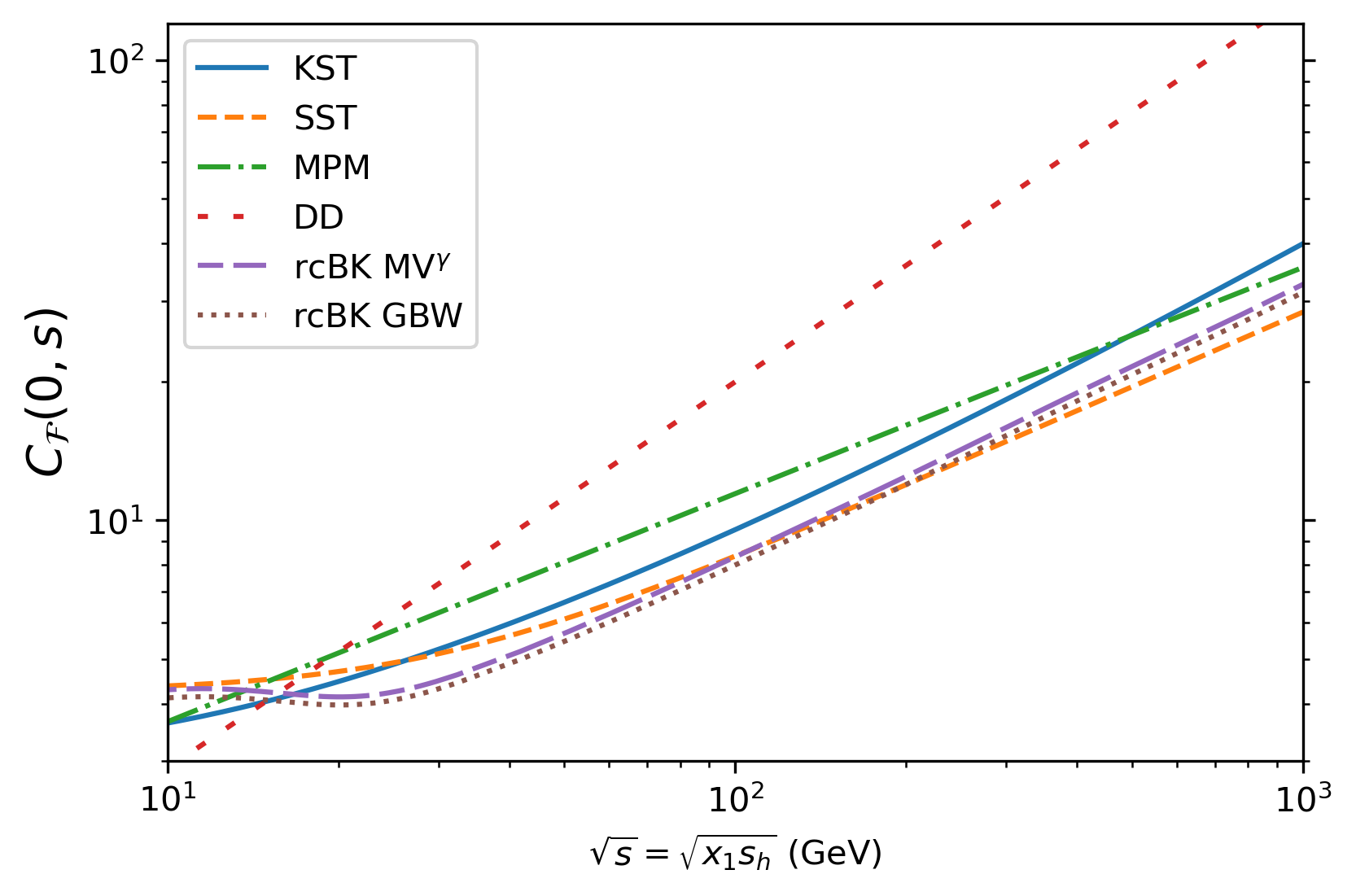}
\caption{Coefficient $\CF(0,s)$ using KST, SST, MPM and DD models, as well as the rcBK  calculations with different AAMQS initial conditions, as a function of $\sqrt{s}$.} 
\label{fig:energy}
\end{figure}


\section{Comparing different models} \label{sec:develop}
\begin{table*}[t]
\centering
\begin{tabular}{c |c |c| c| c | c | c | c | c | c}
\hline\hline
Experiment & $y$ range & $\dpt$ Exp. & $\dpt$ KST &$\dpt$ SST &$\dpt$ MPM &$\dpt$ DD &$\dpt$ GBW &$\dpt$ MV$^\gamma$ \\ 
\hline
LHCb  & $-4.5 < y < -2.5$ & $0.79 \pm 0.12$ & $0.84$ & $0.81$ & $1.01$ & $1.23$ & $0.46$ & $0.48$ \\ 
      & $2    < y < 4$    & $2.05 \pm 0.12$ & $3.90$ & $2.81$ &$3.54$ & $13.23$ & $2.94$ & $3.08$ \\ 
 \hline
ALICE & $-4.46 < y < -2.96$ & $0.68 \pm 0.33$ & $0.70$ & $0.72$ & $0.82$ & $0.86$ & $0.40$ & $0.42$ \\ 
 &      $2.03  < y <  3.53$ & $1.91 \pm 0.42$ & $2.91$ & $2.22$ & $2.85$ & $9.14$ & $2.27$ & $2.37$ \\ 
 \hline
PHENIX & $-2.2 < y <  -1.2$ & $0.43 \pm 0.08$ & $0.57$ & $0.67$ & $0.59$ & $0.48$ & $0.60$ & $0.63$ \\ 
       & $-0.35 < y < 0.35$ & $0.71\pm 0.2$   & $0.74$ & $0.75$ & $0.88$ & $0.97$ & $0.59$ & $0.62$ \\ 
       & $1.2 < y < 2.2$    & $0.43\pm 0.08$  & $0.95$ & $0.86$ & $1.16$ & $1.71$ & $0.75$ & $0.79$ \\ 
 \hline \hline
\end{tabular}
\caption{Values of $\dpt$ from LHCb, ALICE and PHENIX, from Ref. \cite{arleo2020nuclear}, for nuclear broadening of $J/\psi$ at $\sqrt{s_h} = 8.16$ TeV, $\sqrt{s_h} = 5.02$ TeV amd $\sqrt{s_h} = 200$ GeV respectively (labeled `$\dpt$ Exp.'). The remaining columns present the prediction for $\dpt$ using KST, SST, MPM and DD models as well as the rcBK  calculations with different AAMQS initial conditions. All $\dpt$ values are in GeV$^2$.} 
\label{tab:dpt}
\end{table*}

We now proceed to compare the predictions of $\CF(0,s)$ using the models described in the previous section. Fig. \ref{fig:energy} shows the energy dependence of $\CF(0,s)$ as a function of $\sqrt{s}$. Notice that the relevant energy scale is not the hadronic c.m. energy $s_h$, but rather the energy of parton-target system, i.e., $s = x_1 s_h$ (see, for instance, Ref. \cite{raufeisen2004heavy}).

Our main goal is now to evaluate $\dpt$ using these models and compare the results to actual data. Numerical values of $\dpt$ for $J/\psi$ production in pp and pPb collisions were extracted from PHENIX, ALICE and LHCb in Ref. \cite{Arleo:2020rbm} and are presented in the third column of Table \ref{tab:dpt}. The remaining columns of this table present results for $\dpt$ from the different models considered in the previous section and were obtained as described next. As noted in Ref. \cite{raufeisen2003relating}, broadening in $J/\psi$ production is equal to broadening for gluons (assuming final states effects are negligible). We therefore evaluate $\dpt$ using Eq. \eqref{eq:dpt}, with $\mathcal{R}=\mathcal{A}$ and $A = 208$. The value of $s = x_1 s_h$ is obtained using $x_1 = e^y M/\sqrt{s_h}$, with the average value of $y$ in each bin.
We also take into account suppression due to gluon shadowing, as presented in Ref. \cite{kopeliovich2000nonperturbative, raufeisen2004heavy}. While negligible for $x_2 \sim 10^{-2}$, for $x_2 \sim 10^{-5}$ it leads to a reduction of $\CF(0,s)$ by about 1/3. 

In the LHC regime, KST, SST and MPM models lead to reasonable and similar predictions for $\dpt$, while the DD model largely overestimates broadening in the positive $y$ range.  We trace this to the large exponential growth $(s/s_o)^{0.42}$ in Eq. \eqref{eq:cfDD}, meaning that $\CF^{DD}$ grows faster with $\sqrt{s} = \sqrt{x_1 s_h}$ than in the other models, as shown in Fig. \ref{fig:energy}. Although KST, SST and MPM models perform significantly well for backward rapidity (leading to $x_2 \sim 10^{-2}$), we find that SST performs better in the forward rapidity intervals (corresponding to $x_2 \sim 10^{-5}$). In the forward rapidity region, the rcBK results fall in between the SST and MPM ones. The results at backward rapidity are quite different though at TeV energies. The backward rapidity region is associated with a smaller value of $x_1$, which translates to a smaller energy value in the parton-target system. The ALICE and LHCb regime probes $C_{\mathcal{F}}(0,s)$ at $22 < \sqrt{x_1 s_{h}}\,\, (\rm GeV) < 34$, the region where the rcBK results present the lowest values of all models considered (see Fig. \ref{fig:energy}); the PHENIX data probes $C_{\mathcal{F}}(0,s)$ at $\sqrt{x_1 s_{h}}\sim 11$ GeV, region where all models but the DD one coincide.

\begin{figure}[htb]
\centering
\includegraphics[scale=0.6]{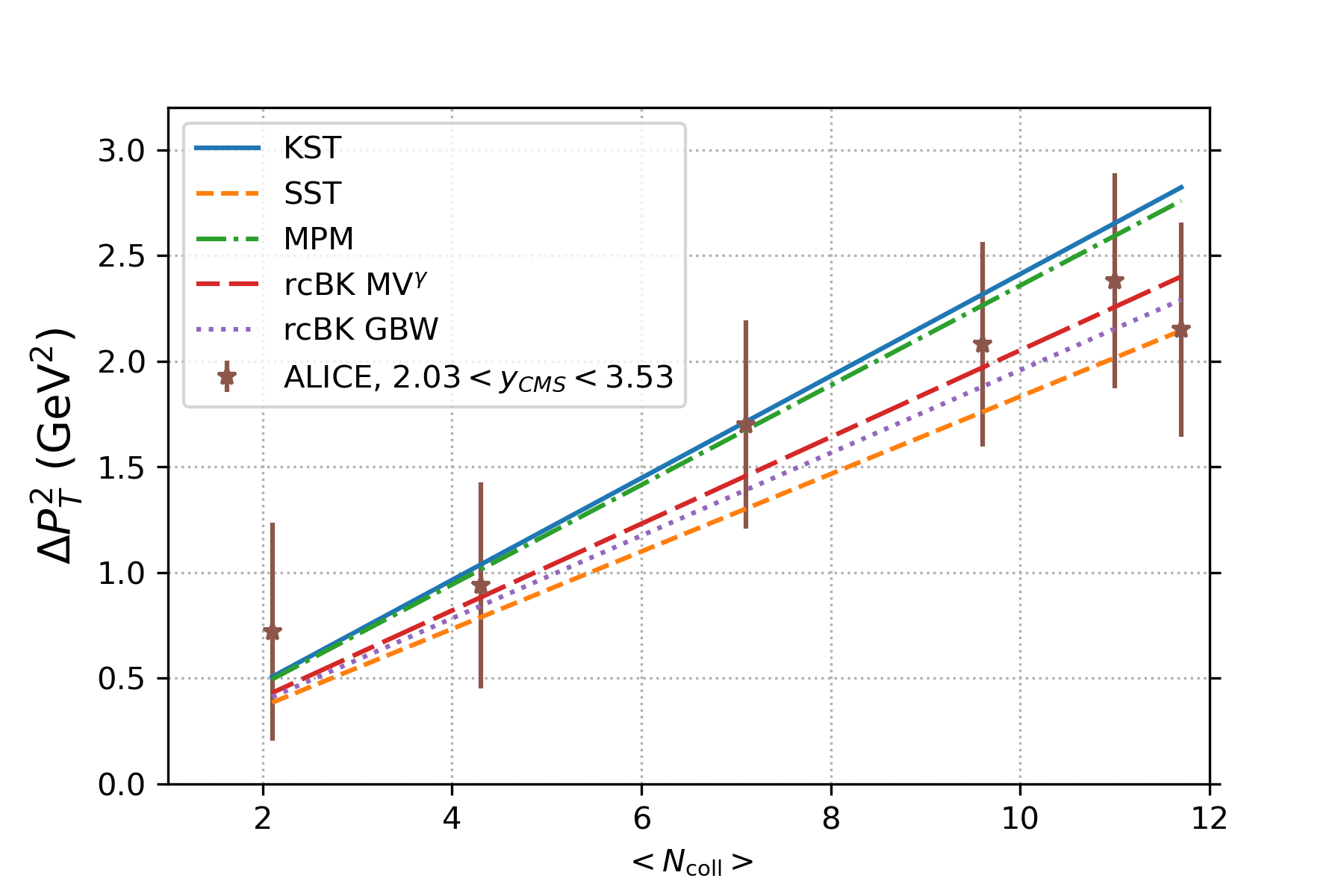}
\caption{Nuclear broadening $\dpt$ as a function of the average number of binary collisions <$N_{\mathrm{coll}}$> in $J/\psi$ production in pPb collisions at $\sqrt{s_h} = 5.02$ TeV. Points represent extracted values from ALICE, with $2.03 < y < 3.53$, from Ref. \cite{ALICE:2015kgk}. The lines represent the prediction for $\dpt$ using KST, SST and MPM models as well as the rcBK  calculations with different AAMQS initial conditions.}
\label{fig:dpt}
\end{figure}

Data for $\dpt$ in $J/\psi$ production in pPb collisions at $\sqrt{s_h} = 5.02$ TeV is also available from ALICE as a function of centrality in Ref. \cite{ALICE:2015kgk}. In each centrality class, we used Eq. \eqref{eq:dpt} with $\langle T_A \rangle$ equal to the average value of the nuclear overlap function $\langle T_{pPb}^{\mathrm{mult}} \rangle$ (see Table 1 in Ref. \cite{ALICE:2015kgk}). Table \ref{tab:dptcentral} presents $\dpt$ using data from ALICE, in the range $2.03 < y < 3.53$, and the predictions using KST, MPM and rcBK models, in each centrality class.  We restrict the analysis just for forward rapidities, where all models are within their range of validity. We checked that the description for backward rapidities are still reasonable.
The same regime is presented in Fig. \ref{fig:dpt}, in which the prediction for $\dpt$ is presented as a function of the average number of binary collisions <$N_{\mathrm{coll}}$> for each model, along with the values extracted from ALICE data. We see that all models predict an increase in broadening with <$N_{\mathrm{coll}}$>, as expected. 
The SST model remains with good adherence to experimental data within the errors; results from the rcBK model with GBW initial conditions are quite similar to the the SST one for all centralities. The other models are not excluded given the experimental uncertainties.

\section{Discussions and Conclusions} \label{sec:conclusions}

\begin{table}[t]
\centering
\begin{tabular}{c |c |c| c | c }
\hline\hline
ZN class & $\dpt$ ALICE & $\dpt$ KST &$\dpt$ MPM &$\dpt$ MV$^\gamma$ \\
\hline
2-10\%   & $2.15 \pm 0.51$ & $2.82$ & $2.75$ & $2.40$  \\
10-20\%  & $2.38 \pm 0.51$ & $2.65$ & $2.59$ & $2.26$  \\
20-40\%  & $2.08 \pm 0.49$ & $2.30$ & $2.25$ & $1.95$  \\
40-60\%  & $1.70 \pm 0.49$ & $1.71$ & $1.67$ & $1.45$  \\
60-80\%  & $0.94 \pm 0.49$ & $1.03$ & $1.01$ & $0.88$  \\
80-100\% & $0.72 \pm 0.52$ & $0.51$ & $0.50$ & $0.43$  \\
 \hline \hline
\end{tabular}
\caption{Values of $\dpt$ in $J/\psi$ production in pPb collisions at $\sqrt{s_h} = 5.02$ TeV from ALICE, with $2.03 < y < 3.53$, from Ref. \cite{ALICE:2015kgk}, as a function of centrality. We also present the prediction for $\dpt$ using KST, MPM and the rcBK  calculation with MV initial conditions. All $\dpt$ values are in GeV$^2$.}
\label{tab:dptcentral}
\end{table}

Before conclusions, we would like to compare the present calculation to other approaches where multiple scattering of the projectile partons is the underlying dynamics. One of them is the Color Glass Condensate formalism. Specifically, the  production of quarkonium at forward rapidities for proton and nuclear targets was addressed in Ref. \cite{Ducloue:2015gfa}. By using a proper treatment of the nuclear geometry, one obtains a nuclear suppression that is in very good agreement to the experimental data. An extension of these investigations has been done in Ref. \cite{Ducloue:2016pqr}, where the mean transverse momentum, $\langle p_T\rangle$,  was analyzed in terms of meson rapidity and collision centrality. It was found an intense variation of $\Delta p_T^2$ as a function of $N_{\mathrm{coll}}$.
The main ingredient of these approaches is the nuclear saturation scale, in which $p_T$-broadening in the saturation regime is roughly expressed as $\langle p_T^2 \rangle \approx Q_{s,A}^2/4$  \cite{Dumitru:2001jn,Schafer:2013mza}. Here, the nuclear saturation scale squared  is $Q_{s,A}^2\sim A^{1/3}s^{\lambda}$ with $\lambda \simeq 0.3$. 

The LHC data for broadening  can also be described through approaches that take into account initial/final-state parton's multiple scattering in the nuclear medium \cite{Kang:2008us,Kang:2012am} or those that include coherent energy loss effects from the incoming and
outgoing partons \cite{Arleo:2013zua,Arleo:2018zjw}. Both predictions are consistent with the ALICE data, with the multiple scattering approach doing a better job for forward rapidities compared to the energy loss model. The collider data for prompt quarkonium production have allowed accurate extraction of the $x$-dependence of the transport coefficient, $\hat{q}(x)$ \cite{Naim:2020bis,Arleo:2020rbm}. In this case, the nuclear broadening is related to $\hat{q}_0=\hat{q}(x=x_0)$ (with $x_0= 10^{-2}$) in the following form:
\begin{eqnarray}
\Delta p_T^2 = \frac{\hat{q}(x)}{N_c}C_{\mathrm{oc,sg}}\Delta L, \quad \hat{q}(x)=  \hat{q}_0 \left(\frac{x_0}{x}  \right)^{\lambda_g},
\end{eqnarray}
where one assumes a power-like behavior for the gluon distribution, $xG(x,\mu^2)\propto x^{-\lambda_g}$ with $\lambda_g\simeq 0.25$. The color factors for quarkonium are  $C_{\mathrm{oc,sg}}=N_c,N_c/2$ for octet and singlet color states, respectively. The nuclear medium length $L_A$ enters into the quantity $\Delta L = L_A-L_p^{\prime}$. The value found for the transport coefficient is $\hat{q}_0\simeq 0.050$ GeV$^2$/fm \cite{Naim:2020bis,Arleo:2020rbm}. This value is somewhat higher than that obtained by a  global analysis of the transport coefficient for cold nuclear matter within the generalized QCD factorization formalism \cite{Ru:2019qvz}, with the best fit giving $\hat{q}_0\simeq 0.02$ GeV$^2$/fm. However, a different functional form for the $\hat{q}(x,\mu^2)$ is employed in Ref. \cite{Ru:2019qvz}.

Concerning the present work, the results obtained by using the KST model are consistent with those predicted in Ref. \cite{Kopeliovich:2017jpy}, namely $\Delta p_T^2 (-4.46 \leq y \leq -2.96)=0.35 $, $\Delta p_T^2 (-1.37\leq y\leq -0.43)=0.73 $ and $\Delta p_T^2 (2.03 \leq y \leq  3.53)= 2.27$ GeV$^2$ at $\sqrt{s}= 5.02$ TeV.  We find that, although the models KST, SST and MPM predict values of $\dpt$ that agree reasonably well with experiment, the SST model leads to better results, specially for forward rapidity. Results with rcBK evolved dipole cross sections are quite close to the SST model at forward rapidities and present the smaller values of $p_T$-broadening at backward rapidities. The DD model predicts stronger broadening than the others. The reason is the high value for the hard Pomeron intercept, $\alpha_{I\!\!P}(0)=1.42$. The dependence on centrality measured by ALICE is correctly described by the different models. Although some of those models include gluon saturation corrections, the experimental results are not able to clearly disentangle the different QCD dynamics embedded in the considered models. This is, in fact, a current challenge, and intense work is ongoing in order to find different ways of identifying the saturated gluon state \cite{Morreale:2021pnn}.
\section*{Acknowledgements} \label{sec:acknowledgements}
This work was financed by the Brazilian funding agency CNPq and in part by CAPES - Finance Code 001.
The work of AVG has been supported by FAPESP through grants 17/05685-2 and 21/04924-9.

\bibliography{fgbenbib}


\end{document}